\documentclass[twocolumn,showpacs,preprintnumbers,amsmath,amssymb,prb]{revtex4}
\usepackage[dvips]{color,graphics,epsfig,rotating}
\usepackage{graphicx}
\usepackage{dcolumn}
\usepackage{bm}

\begin{document}

\title{Density functional study of FeS, FeSe and FeTe:
Electronic structure, magnetism, phonons and superconductivity}

\author{Alaska Subedi}
\affiliation{Department of Physics and Astronomy,
University of Tennessee, Knoxville, TN 37996}
\affiliation{Materials Science and Technology Division,
Oak Ridge National Laboratory, Oak Ridge, Tennessee 37831-6114} 

\author{Lijun Zhang}
\affiliation{Materials Science and Technology Division,
Oak Ridge National Laboratory, Oak Ridge, Tennessee 37831-6114} 

\author{D.J. Singh}
\affiliation{Materials Science and Technology Division,
Oak Ridge National Laboratory, Oak Ridge, Tennessee 37831-6114} 

\author{M.H. Du}
\affiliation{Materials Science and Technology Division,
Oak Ridge National Laboratory, Oak Ridge, Tennessee 37831-6114} 

\date{\today} 

\begin{abstract}
We report density functional calculations of the electronic structure,
Fermi surface, phonon spectrum, magnetism and electron-phonon
coupling for the superconducting phase FeSe, as well as the related
compounds FeS and FeTe.
We find that the Fermi surface structure of these
compounds is very similar to that of the
Fe-As based superconductors, with
cylindrical electron sections at the zone corner,
cylindrical hole surface
sections, and depending on the compound, other small hole sections
at the zone center.
As in the Fe-As based materials, these surfaces are separated
by a 2D nesting
vector at ($\pi$,$\pi$). The density of states, nesting
and Fermi surface size increase going from FeSe to FeTe.
Both FeSe and FeTe show spin density wave ground states, while FeS
is close to an instability. In a scenario where superconductivity
is mediated by spin fluctuations at the SDW nesting vector, the
strongest superconductor in this series would be doped FeTe.
\end{abstract}


\maketitle

\section{introduction}

Superconductivity was recently reported in 
$\alpha$-FeSe$_{1-x}$, with critical
temperature $T_c \sim$ 8K. \cite{hsu}
This is of interest both because of the 
fact that Fe containing superconductors are unusual and because
this material shares in common square planar sheets of tetrahedrally
coordinated Fe with the Fe-As based high temperature
superconductors. \cite{kamihara,ren,ren2,cwang,sefat,kito}
$\alpha$-FeSe occurs in the PbO structure. This consists, as mentioned,
of Fe square planar sheets, with Se atoms forming distorted
tetrahedra around the
Fe very similar to the structure of the FeAs planes in
LaFeAsO, BaFe$_2$As$_2$ and LiFeAs, which are prototypes of the
known families of Fe-As based high-T$_c$ superconductors.
\cite{rotter1,rotter2,xcwang}

Since the reported $T_c$ = 8K of doped $\alpha$-FeSe is modest,
it is important first of all to establish the relationship between
this material and the Fe-As based superconductors. We note
that LaNiPO is also a superconductor and shares the crystal structure
of LaFeAsO,
\cite{watanabe,tegel}
but that it is apparently quite different electronically
and can be understood in terms of standard 
electron-phonon theory,\cite{subedi}
unlike the Fe-As based phases. \cite{boeri,mazin}

Here we report density functional calculations that show $\alpha$-FeSe
and the other known Fe based chalcogenides in this structure to be
very similar to that of the Fe-As based superconductors. \cite{singh-du}
In particular the Fermi surface consists of small
heavy hole cylinders near the zone center and lighter compensating
electron cylinders around the zone corner.
We show that
the stoichiometric compounds are either very close to a spin density
wave instability (FeS) or have an
itinerant spin density wave instability without doping (FeSe and
FeTe) similar to the Fe-As superconductors. \cite{cruz}
We predict that this itinerant
nesting driven magnetic state is strongest in FeTe and in addition that
FeTe has the largest Fermi surface of the three compounds.
Calculations of the electron-phonon coupling show that doped FeSe is
not an electron-phonon superconductor, similar to what was
found for the Fe-As phases.
Within a spin-fluctuation driven picture of superconductivity the results
indicate that FeTe with doping is a likely higher temperature superconductor.

\section{first principles methods and structure}

Our calculations of the electronic structure and magnetic properties
were performed within the local density approximation with the
general potential linearized augmented planewave (LAPW) method,
\cite{singh-book}
including local orbitals, \cite{singh-lo}
similar to our previous calculations for the Fe-As based superconductors.
\cite{singh-du,mazin,singh-ba}
We used LAPW spheres of radius 2.1 $a_0$ for Fe, Se and Te
and 1.9 $a_0$ for S.
These compounds occur in a simple tetragonal structure with one internal
parameter, $z_X$ corresponding to the chalcogen height above
the Fe square plane.
The experimental lattice parameters
\cite{hsu,lennie,finlayson}
were employed and we relaxed the chalcogen height via energy minimization.
The structural parameters used and some results are presented
in Table \ref{tab-struct}.
The electron-phonon coupling and phonon dispersions were on the
other hand obtained using linear response, again with the
experimental lattice parameters, with the Quantum Espresso
code \cite{qe,qenote}
within the generalized gradient approximation of
Perdew, Burke and Ernzerhof \cite{pbe} as described for
LaFeAsO and LaNiPO. \cite{subedi,mazin}

\begin{table}
\caption{Structural parameters and magnetic properties of
PbO-structure Fe$X$.
The lattice parameters are from experimental data, while the internal chalcogen
structural parameter, $z_X$ is from LDA structure minimization. $m_{SDW}$ is the
spin moment within the Fe LAPW sphere (radius 2.1 $a_0$) for the SDW
state and $E_{SDW}$ is the energy per Fe of this state relative to the
non-spin-polarized state in meV/Fe.
$N(E_F)$ is the density of states at the Fermi energy in the non-spin-polarized
band structure in eV$^{-1}$ on a per Fe both spins basis.}
\label{tab-struct}
\begin{tabular}{lcccccc}
\hline
  & ~~$a$(\AA)~~ & ~~$c$(\AA)~~ & ~~$z_X$~~ &
                             $N(E_F)$ & $m_{SDW}(\mu_B)$ & $E_{SDW}$ \\
\hline
FeS  & 3.6735 & 5.0328 & 0.2243 & 1.35 & 0.00 & 0 \\
FeSe & 3.765  & 5.518  & 0.2343 & 0.95 & 0.65 & 5 \\
FeTe & 3.8215 & 6.2695 & 0.2496 & 1.83 & 1.28 & 47 \\
\hline
\end{tabular}
\end{table}

\section{electronic structure, phonons and magnetism}

\begin{figure}[tbp]
\includegraphics[height=3.4in,angle=270]{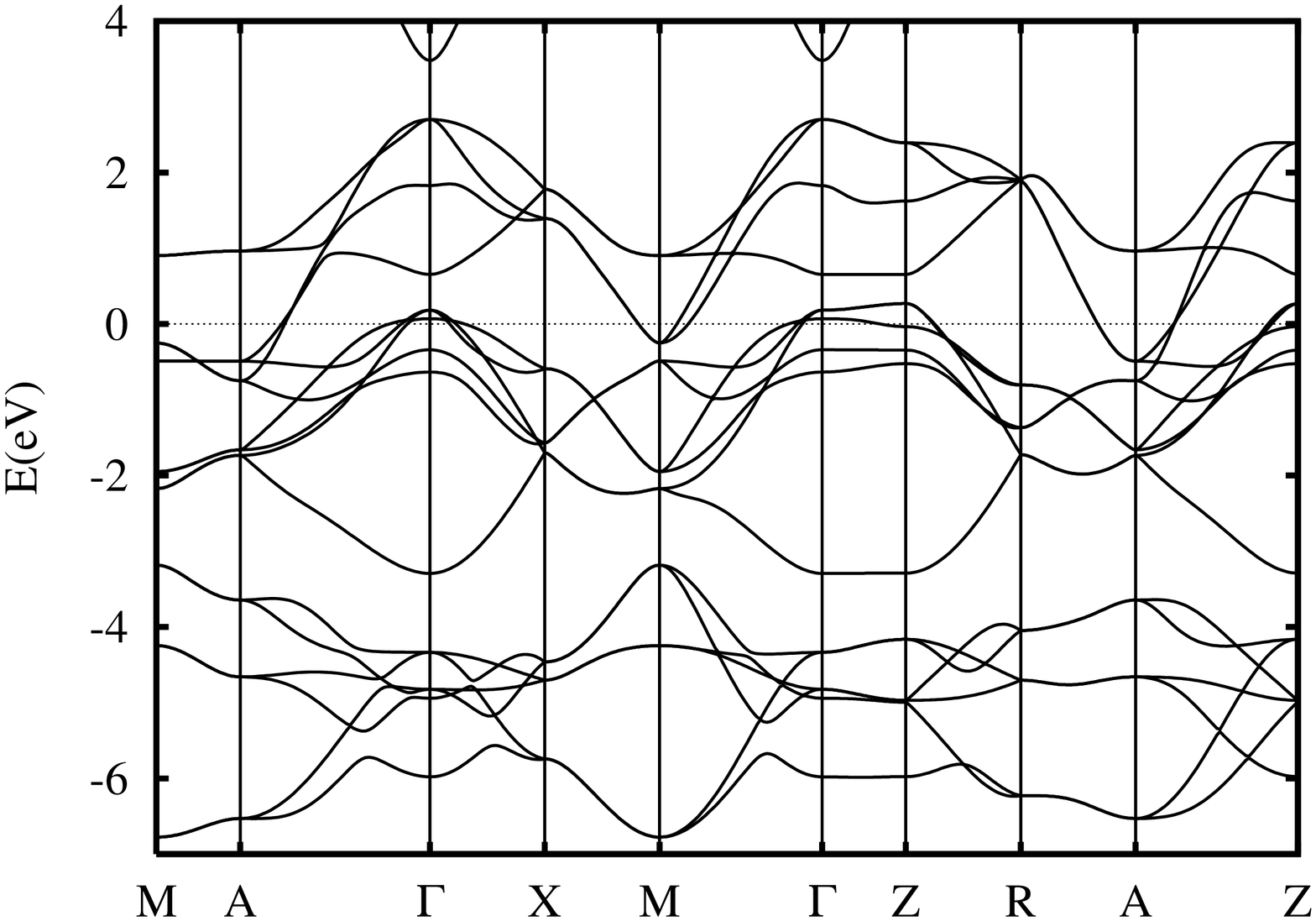}
\includegraphics[height=3.4in,angle=270]{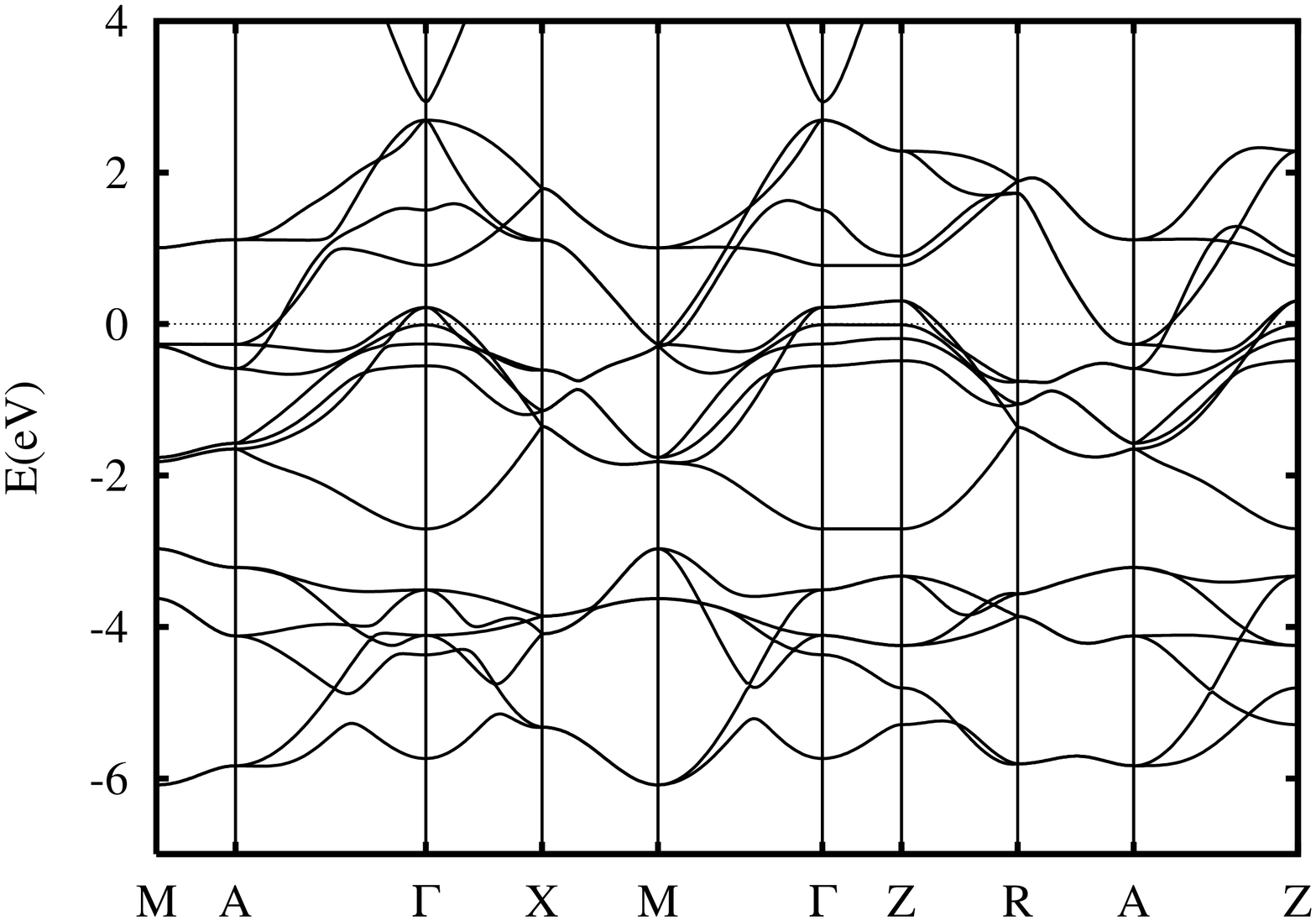}
\includegraphics[height=3.4in,angle=270]{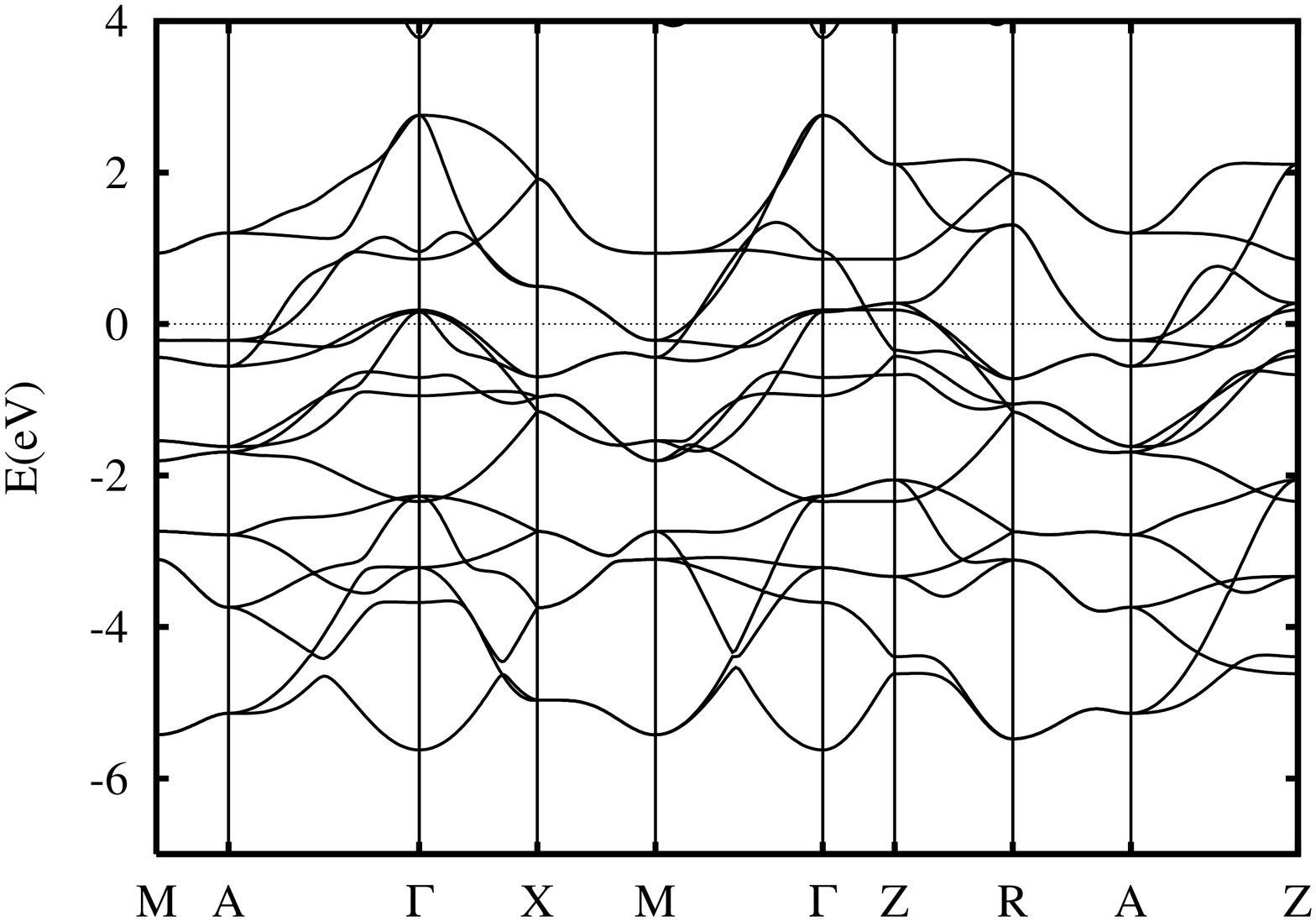}
\caption{Band structures of FeS (top), FeSe (middle) and FeTe (bottom)
from non-spin-polarized calculations with the LDA relaxed $X$
heights.}
\label{fig-bands}
\end{figure}

\begin{figure}[tbp]
\includegraphics[height=3.4in,angle=270]{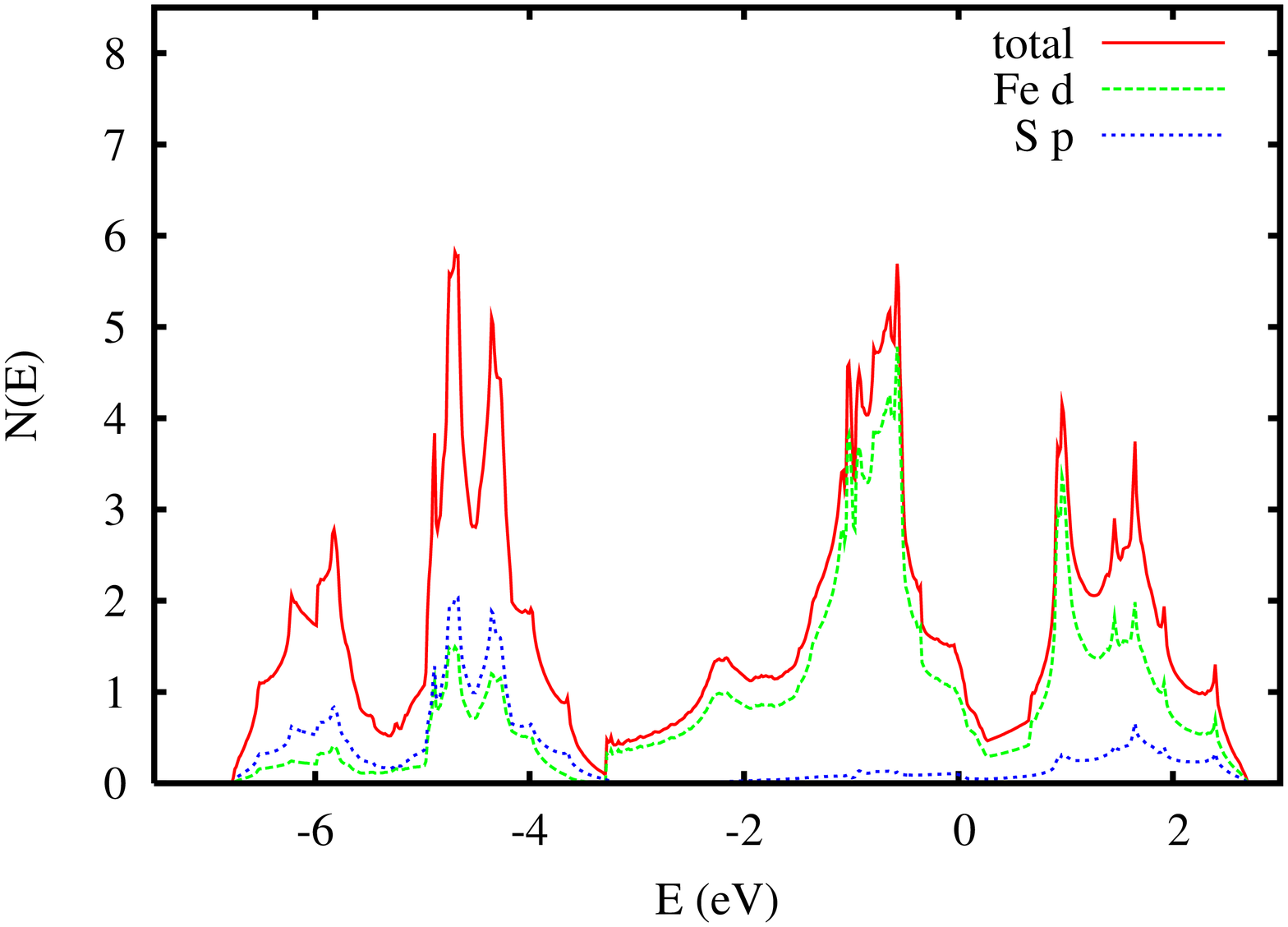}
\includegraphics[height=3.4in,angle=270]{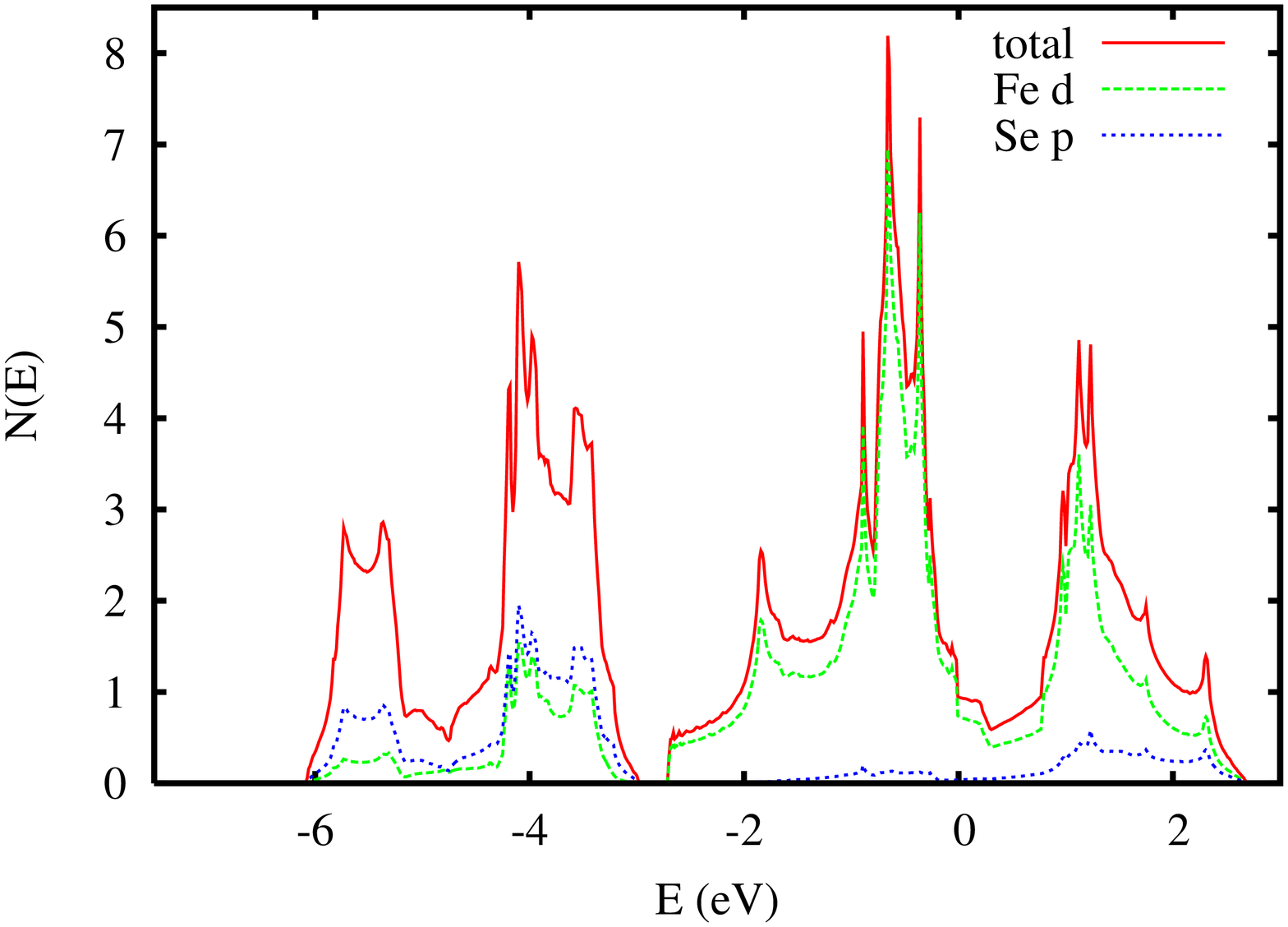}
\includegraphics[height=3.4in,angle=270]{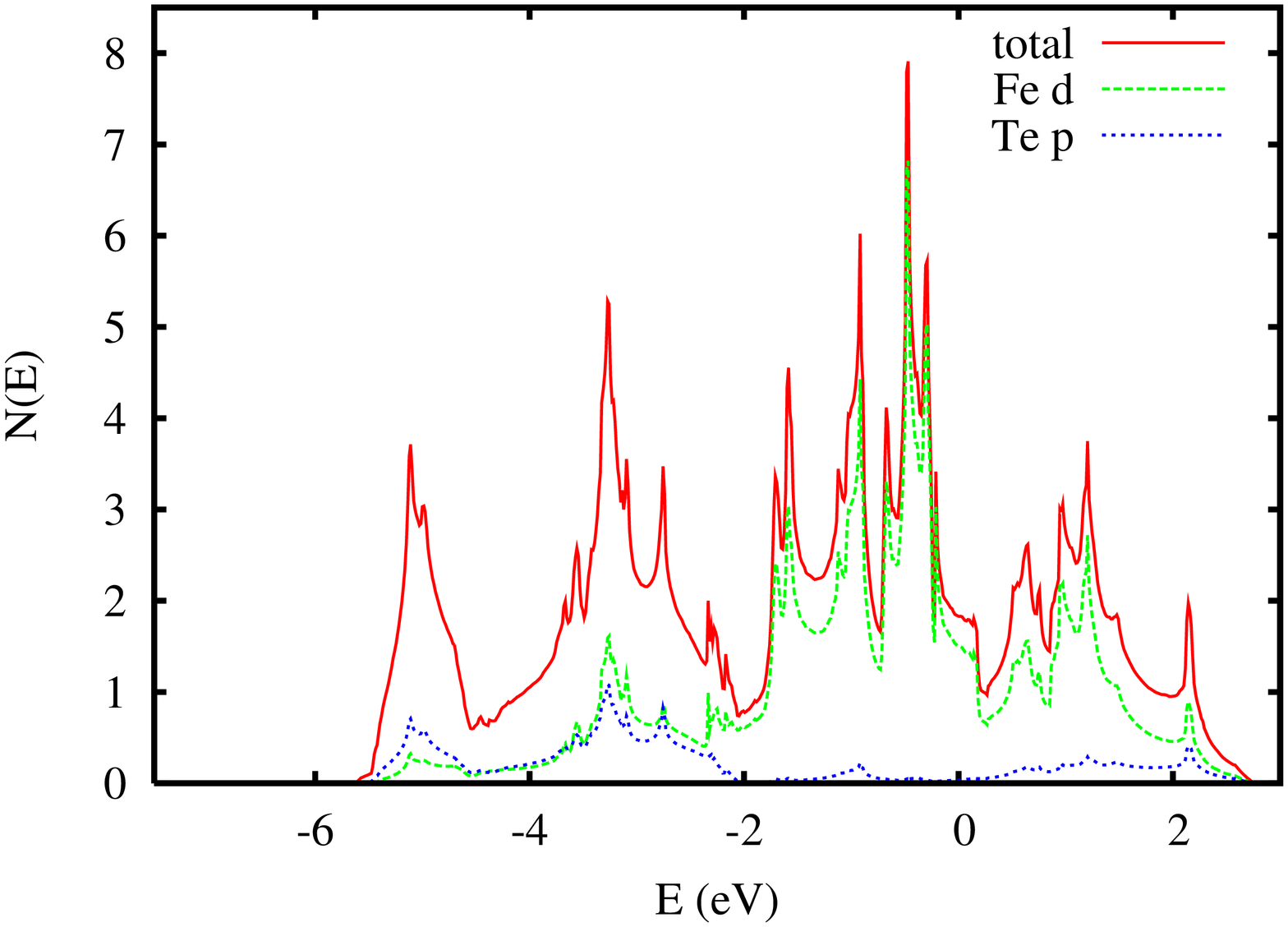}
\caption{(color online)
Electronic DOS and projection
onto the LAPW Fe and chalcogen spheres indicating the Fe $d$ and
chalcogen $p$ contributions for
FeS (top), FeSe (middle) and FeTe (bottom)
as in Fig. \ref{fig-bands}.}
\label{fig-dos}
\end{figure}

\begin{figure}[tbp]
\includegraphics[width=3.4in]{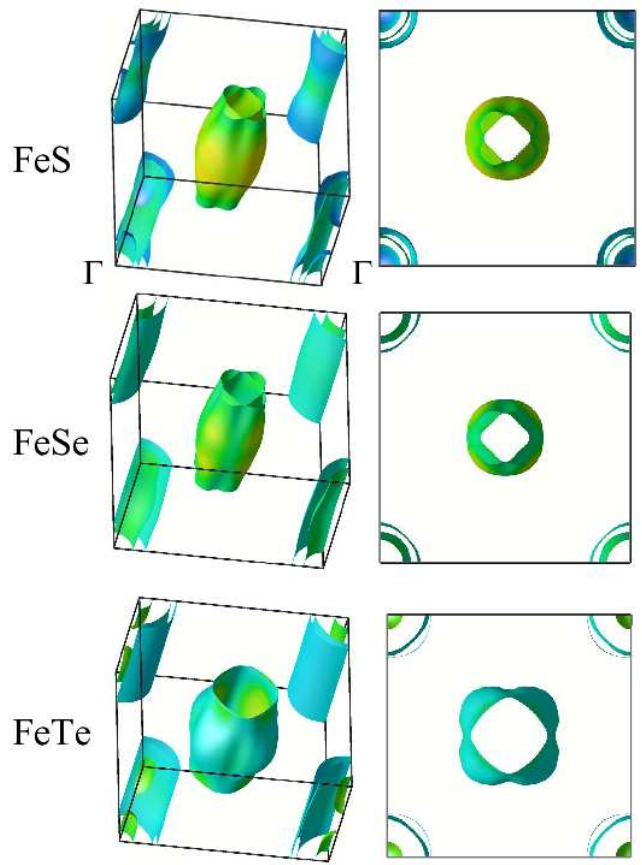}
\caption{(color online) LDA Fermi surface of FeS, FeSe and FeTe
from non-spin-polarized calculations with the LDA relaxed $X$
heights. The corners are $\Gamma$ points.}
\label{fig-fermi}
\end{figure}

\begin{figure}[tbp]
\includegraphics[width=3.4in]{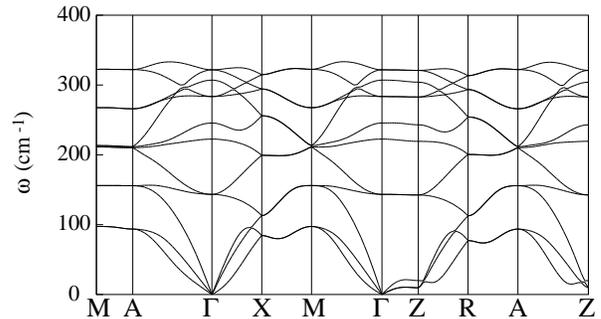}
\caption{Calculated GGA phonon dispersions of non-spin-polarized FeSe.
}
\label{fig-phonon}
\end{figure}

Our main results for the electronic structure are given in
Figs. \ref{fig-bands}, \ref{fig-dos} and \ref{fig-fermi},
which show the non-spin-polarized
band structures, electronic densities of states and
Fermi surfaces of FeS, FeSe and FeTe.
The calculated values of $N(E_F)$ are given in Table \ref{tab-struct}.
The phonon dispersions of FeSe
are shown in Fig. \ref{fig-phonon}.
One interesting feature of the phonon dispersion is that they have
little dispersion in the $k_z$ and e.g. are very
flat along the tetragonal $\Gamma$-$Z$ direction. This presumably reflects
anion-anion repulsion, which leads to long bonds between the FeSe layers.
the result is that there may be an easy cleavage plane between the
Se ions, which may facilitate preparation of clean surfaces for experiments
such as photoelectron spectroscopy.

\begin{figure}[tbp]
\includegraphics[width=3.4in]{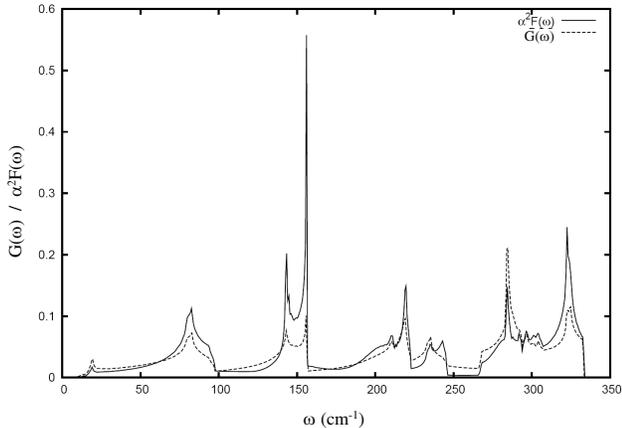}
\caption{Calculated GGA phonon density of state $G(\omega)$ and
electron-phonon spectral function $\alpha^2 F(\omega)$ for FeSe.
}
\label{fig-a2f}
\end{figure}

The phonon density of states, $G(\omega)$ and electron-phonon
spectral function,
$\alpha^2 F(\omega)$ are given in Fig. \ref{fig-a2f}.
The electron-phonon coupling constant for FeSe
as obtained in linear response is
$\lambda$ = 0.17 with $\omega_{log}$=113 cm$^{-1}$.
No superconductivity at any temperature even approaching 1K
results with these values within standard electron-phonon theory
even if very low values of the Coulomb parameter, e.g. $\mu^*=0.10$ are used in
the Allen-Dynes equation.
This is similar to what was found previously for LaFeAsO.
\cite{boeri,mazin}
Therefore we conclude that FeSe is not a conventional electron-phonon
superconductor.

Turning to the electronic structure we find a strong qualitative
similarity between these materials and the FeAs-based superconductors.
In particular, we find these to be low carrier density metals, with
high density of states. This arises from band structures that
are closely related to those of the FeAs materials.
The chalcogen $p$ states lie well below the Fermi level and are
only modestly hybridized with the Fe $d$ states as may be
seen from the projected DOS (Fig. \ref{fig-dos}).
Thus the electronic structure near the Fermi energy derives from
metallic Fe$^{2+}$ layers, with direct Fe-Fe interactions. These
are embedded
inside a largely ionic background which imposes a competing tetrahedral
crystal field.
As in the Fe-As based materials, there is a pseudogap at an electron
count of 6 $d$ electrons per Fe, and $E_F$ lies near the bottom of this
pseudogap. We emphasize that this is not the position of a tetrahedral
crystal field gap, which would be at 4 electrons, and emphasizes the
fact that Fe chalcogen hybridization is not strong compared with the Fe-Fe
interactions. This explains the similarity of the electronic structure
to that of the FeAs-based materials, which were also found to be substantially
ionic in similar calculations. \cite{singh-du} 

These band structures yield two intersecting elliptical cylindrical
electron Fermi surfaces at the zone corner in all three materials. These
are compensated by lower velocity hole sections at the zone center --
two concentric hole cylinders, and in the case of FeS and FeTe a small
closed hole section inside the inner cylinder.
This is qualitatively very similar to the FeAs-based materials.
It is important to note that cylinders at the zone center and zone
corner, if they are the same size, would yield strong nesting peaked
at the 2D ($\pi$,$\pi$) point. This will lead in general to enhanced
spin fluctuations at the nesting vector, and if sufficiently strong
will cause a spin density wave. In fact, an ordered SDW was
found both in first principles calculations and in experimental
studies for LaFeAsO and many other undoped FeAs-based compounds.
\cite{mazin,cruz,dong,chen,zhao,huang,goldman,nakai}
There is a clear competition between the SDW and the superconducting
state in that superconducting samples generally do not show the SDW,
while samples with the SDW transition, generally do not show
superconductivity.
The ground state is an antiferromagnetic cell doubled along the
[11] in plane direction to yield linear chains of nearest neighbor
like spin Fe atoms arranged antiferromagnetically,
although the values of the moments are dependent on details,
especially the As height above the Fe plane.
\cite{yildirim,ishibashi,yin,mazin-2,vildosola,yildirim-2}

\section{discussion}

Turning to the trends, the size of the pseudogap is approximately the
same in FeSe and FeS, but is significantly smaller in FeTe. Specifically,
there is a greater overlap between the hole and electron bands in the latter
compound. This leads to larger Fermi surfaces. The value of
$N(E_F)$=1.83 eV$^{-1}$ is
also highest in this compound, although it is still lower than the
2.62 eV$^{-1}$ that is obtained for LaFeAsO by the same approach.
We used a supercell approach to investigate the SDW with the chalcogen
heights fixed to the values calculated by non-spin-polarized energy
minimization. We find instabilities for FeSe and FeTe, but not
for FeS. The spin moments and the energy of the SDW relative to the
non-spin-polarized state are given in Table \ref{tab-struct}.
As may be seen the SDW is considerably stronger in FeTe than in FeSe,
with an energy gain of 47 meV/Fe and a spin-moment of 1.3 $\mu_B$.
The corresponding values for LaFeAsO, calculated in the same way
are $E_{SDW}$=11 meV/Fe and $m_{SDW}$ = 1.0 $\mu_B$ in a 2.1 $a_0$ Fe sphere.
Besides the SDW we find a borderline ferromagnetic tendency in FeTe,
without doping when the SDW
is not allowed, similar to LaFeAsO. \cite{singh-du}
We do not find ferromagnetic instabilities in either FeSe or FeS, consistent
with the lower values of $N(E_F)$ in those materials.
The sensitivity of the moment size to the ordering pattern
underscores that fact that
these are itinerant magnetic systems in the LDA, as opposed to local moment
magnets.
This means that magnetic
ordering is driven by electrons at and near the Fermi surface.
On the other hand this is not to say that spin fluctuations are weak
in the paramagnetic state above the SDW ordering temperature or the
paramagnetic state
as realized by doping. In fact, as noted, several authors have
found that the As height in the Fe-As compounds is strongly coupled to
magnetism and so strong spin fluctuations in the paramagnetic state
would help rationalize the underestimated As height in
non-spin-polarized LDA calculations.
\cite{yildirim,ishibashi,yin,mazin-2,yildirim-2}
In fact, there is evidence for strong spin fluctuations in the normal
state of the FeAs compounds,
e.g. from temperature dependent resistivity data indicating strong
scattering above the SDW ordering temperature (note a drop in resistivity
below the SDW ordering temperature even though the carrier density 
is strongly reduced), \cite{mcguire}
as well as from spectroscopy. \cite{mannella}
The size of the effects observed implies that these fluctuations
should have large amplitudes and therefore unlike
the SDW should be rather diffuse
in ${\bf q}$-space, which might make them hard to directly observe.
\cite{sf-note}
Furthermore, the strong transport signatures especially the enhanced
resistivity above the ordering temperature imply substantial coupling
between spin flucuations and electrons at the Fermi energy, which is
indicative of the itinerant nature of the magnetism.

We emphasize also that our results in Table \ref{tab-struct} are at the
LDA relaxed atomic positions for the non-spin-polarized states. These
systems become more magnetic as the chalcogen height is raised.
In the case of FeS a recent refinement is available, and
gives $z_{\rm S}$=0.2602. \cite{lennie}
This puts the S ions 0.18 \AA~ higher than in the LDA structure.
With this value we find a stable
SDW for FeS, with a moment of 1.2 $\mu_B$/Fe.
This is the same trend as in the Fe-As based superconducting materials.
The result
may be taken as an indication that in fact FeS may have a spin density
wave as well and at least that there will be strong spin fluctuations in 
FeS as well as the other PbO-structure Fe chalcogenides.

As mentioned, cylindrical Fermi surface sections of equal volume will
be nested with nesting vector equal to the separation of the centers
of these cylinders. This nesting can be reduced by imperfect matches
in shape, three dimensionality, and size mismatch. Size mismatch
can arise both from additional Fermi surface sections, such as the
extra small hole sections obtained in FeS and FeTe, but not FeSe,
or from doping. In particular, electron doping will reduce the size of
the hole sections and increase the size of the electron sections consistent
with Luttinger's theorem.

The mechanism for superconductivity in the Fe-As based phases has not
yet been established. Nonetheless, there are indications that magnetism
is associated with superconductivity. These include the modest electron
phonon couplings in the materials, the proximity to magnetism and
the phase diagrams which show an association between the SDW and
superconductivity. We discuss our results within a general spin fluctuation
mediated framework. \cite{moriya,monthoux,mazin,kuroki}
In general an itinerant SDW instability arises from a divergence of the
real part of the susceptibility $\chi({\bf q})$
at a specific wavevector ${\bf q}$.
Superconducting pairing is also associated with the real part of
$\chi({\bf q})$ through an integral over the Fermi surface. 
$\chi({\bf q})$ for ${\bf q}$ connecting different parts of the
Fermi surface can contribute to pairing.
As a result, when the Fermi surfaces are small and disconnected as in
these Fe based materials, the fluctuations associated with nesting
will provide substantial interband pairing between the electron
and hole sections, but will not provide substantial intraband pairing.
\cite{mazin,kuroki}
As the nesting is reduced, e.g. by doping, 
$\chi({\bf q})$
will spread out while the peak value
will be reduced, consistent with the destruction of the SDW.
For circular cylinders of radii differing by $\delta q$,
$\chi({\bf q})$
will show a plateau of high
$\chi({\bf q})$
around the nesting vector with width $2 \delta q$
(this will persist until the radii differ by a factor of two
at which point the center of the plataeu will dip).
This means that the two cylinders will still be connected by spin
fluctuations associated with their now reduced nesting, and therefore
even though the SDW will be suppressed by the reduction in the maximum
value of $\chi$ the associated spin fluctuations can still provide
superconducting pairing.
We emphasize that in this general framework the same parts of the
Fermi surface are affected by the SDW and by the superconducting pairing.
Thus these two Fermi surface instabilities compete for the same
Fermi surface and therefore that there should be at best little coexistence
of these two orders.

\section{summary and conclusions}

We report electronic structures, magnetic properties and electron-phonon
calculations for Fe$X$, $X$=S,Se,Te. We find strong similarities to the
Fe-As based superconductors, reflecting the ionic nature of the As and
chalcogen atoms in these compounds. As in the arsenides, we find that
the electron-phonon coupling cannot explain the superconductivity,
and furthermore that these compounds display itinerant magnetism.
These results imply a similar superconducting nature for the Fe-As
phases and FeSe.

The trend that we find in going from FeSe to FeTe is interesting in this
context. In particular we find quite cylindrical Fermi surfaces and
an SDW instability in both compounds. However, the strength of the SDW
is substantially higher in FeTe as is the size of the Fermi surface.
Within the general framework discussed above, FeTe would be expected to have
stronger pairing, and therefore higher $T_c$ than FeSe assuming that
the same mechanism applies in both materials, that both materials can
be chemically doped to the optimum carrier density and that
competing instabilities do not prevent superconductivity in that case.
It will be of interest to experimentally
probe the similarities of FeSe with those of the Fe-As phases
and to search for superconductivity in doped FeTe and in the
alloy Fe(Se,Te).

\acknowledgements
We are grateful for helpful discussions with I.I. Mazin,
D.G. Mandrus and B.C. Sales.
This work was supported by the Department of Energy, Division of
Materials Sciences and Engineering.

\section{note added}
Mizuguchi and co-workers \cite{mizuguchi} recently reported observation
of superconductivity at 27K in FeSe under pressure. This is consistent
with the conclusion here regarding the relationship between FeSe and
the Fe-As based materials.


\begin{references}

\bibitem{hsu}
F.C. Hsu, J.Y. Luo, K.W. Yeh, T.K. Chen, T.W. Huang, P.M. Wu, Y.C. Lee,
Y.L. Huang, Y.Y. Chu, D.C. Yan, and M.K. Wu,
arXiv:0807.2369 (2008).

\bibitem{kamihara}
Y. Kamihara, T. Watanabe, M. Hirano, and H. Hosono,
J. Am. Chem. Soc. {\bf 130}, 3296 (2008).

\bibitem{ren}
Z.A. Ren, W. Lu, J. Yang, W. Yi, X.L. Shen, Z.C. Li, G.C. Che, X.L. Dong,
L.L. Sun, F. Zhou, and Z.X. Zhao,
Chin. Phys. Lett. {\bf 25}, 2215 (2008).

\bibitem{ren2}
Z.A. Ren, G.C. Che, X.L. Dong, J. Yang, W. Lu, W. Yi, X.L. Shen, Z.C. Li,
L.L. Sun, F. Zhou, and Z.X. Zhao,
Europhys. Lett. {\bf 83}, 17002 (2008).

\bibitem{cwang}
C. Wang, L. Li, S. Chi, Z. Zhu, Z. Ren, Y. Li, Y. Wang,
X. Lin, Y. Luo, S. Jiang, X. Xu, G. Cao, and Z. Xu,
arXiv:0804.4290 (2008).

\bibitem{sefat}
A.S. Sefat, M.A. McGuire, B.C. Sales, R. Jin, J.Y. Howe, and D. Mandrus,
Phys. Rev. B {\bf 77}, 174503 (2008).

\bibitem{kito}
H. Kito, H. Eisaki, and A. Iyo,
J. Phys. Soc. Japan {\bf 77}, 063707 (2008).

\bibitem{rotter1}
M. Rotter, M. Tegel, D. Johrendt, I. Schellenberg, W. Hermes, and
R. Pottgen,
arXiv:0805.4021 (2008).

\bibitem{rotter2}
M. Rotter, M. Tegel, and D. Johrendt, arXiv:0805.4630 (2008).

\bibitem{xcwang}
X.C. Wang, Q.Q. Liu, Y.X. Lu, W.B. Gao, L.X. Yang, R.C. Yu, F.Y. Li,
and C.Q. Jin,
arXiv:0806.1687 (2008).

\bibitem{watanabe}
T. Watanabe, H. Yanagi, T. Kamiya, Y. Kamihara, H. Hiramatsu, M. Hirano,
and H. Hosono,
Inorg. Chem. {\bf 46}, 7719 (2007).

\bibitem{tegel}
M. Tegel, D. Bichler, and D. Jorendt,
Solid State Sciences {\bf 10}, 193 (2008).

\bibitem{subedi}
A. Subedi, D.J. Singh, and M.H. Du,
arXiv:0806.3785 (2008).

\bibitem{boeri}
L. Boeri, O.V. Dolgov, and A.A. Golubov,
Phys. Rev. Lett. {\bf 101}, 026403 (2008).

\bibitem{mazin}
I.I. Mazin, D.J. Singh, M.D. Johannes, and M.H. Du,
Phys. Rev. Lett. {\bf 101}, 057003 (2008).

\bibitem{singh-du}
D.J. Singh and M.H. Du, Phys. Rev. Lett. {\bf 100}, 237003 (2008).

\bibitem{cruz}
C. de la Cruz, Q. Huang, J.W. Lynn, J. Li, W. Ratcliff II, 
J.L. Zaretsky, H.A. Mook, G.F. Chen, J.L. Luo, N.L. Wang, and P. Dai,
Nature {\bf 443}, 899 (2008).

\bibitem{singh-book}
D.J. Singh and L. Nordstrom, {\em Planewaves, Pseudopotentials and the
LAPW Method, 2nd. Ed.} (Springer, Berlin, 2006).

\bibitem{singh-lo}
D. Singh, Phys. Rev. B {\bf 43}, 6388 (1991).

\bibitem{singh-ba}
D.J. Singh, arXiv:0807.2643 (2008).

\bibitem{lennie}
A.R. Lennie, S.A.T. Redfern, P.F. Schofield, and D.J. Vaughan,
Mineralogical Magazine {\bf 59}, 677 (1995).

\bibitem{finlayson}
D.M. Finlayson, D. Greig, J.P. Llewellyn, and T. Smith,
Proc. Phys. Soc. B {\bf 69}, 860 (1956).

\bibitem{qe}
S. Baroni, A. dal Corso, S. de Gironcoli, P. Gianozzi,
C. Cavazzoni, G. Ballabio, S. Scandolo, G. Chiarotti, P. Focher,
A. Pasquarello, et al., http://www.quantum-espresso.org.

\bibitem{qenote}
Calculations of the phonon spectra and electron-phonon
coupling were done using ultrasoft pseudpootentials
for Fe, and norm conserving pseudopotentials for Se. We used
a 50 Ry basis set cutoff and a high 500 Ry cutoff for the charge density
expansion. This was needed in order to obtain convergence in the linear
response calculations.

\bibitem{pbe}
J.P. Perdew, K. Burke, and M. Ernzerhof,
Phys. Rev. Lett. {\bf 77}, 3865 (1996).  

\bibitem{dong}
J. Dong, H.J. Zhang, G. Xu, Z. Li, G. Li, W.Z. Hu, D. Wu, G.F. Chen,
X. Dai, J.L. Luo, Z. Fang, and N.L. Wang,
Europhys. Lett. {\bf 83}, 27006 (2008).

\bibitem{chen}
G.F. Chen, Z. Li, D. Wu, G. Li, W.Z. Hu, J. Dong, P. Zheng, J.L. Luo,
and N.L. Wang,
Phys. Rev. Lett. {\bf 100}, 247002 (2008).

\bibitem{zhao}
J. Zhao, Q. Huang, C. de la Cruz, S. Li, J.W. Lynn, Y. Chen, M.A. Green,
G.F. Chen, G. Li, Z. Li, J.L. Luo, N.L. Wang, and P. Dai,
arXiv:0806.2528 (2008).

\bibitem{huang}
Q. Huang, Y. Qiu, W. Bao, J.W. Lynn, M.A. Green, Y.C. Gasparovic,
T. Wu, G. Wu, and X.H. Chen,
arXiv:0806.2776 (2008).

\bibitem{goldman}
A.I. Goldman, D.N. Argyriou, B. Ouladdiaf, T. Chatterji, A. Kreyssig,
S. Nandi, N. Ni, S.L. Budko, P.C. Canfield, and R.J. McQueeney,
arXiv:0807.1525 (2008).

\bibitem{nakai}
Y. Nakai, K. Ishida, Y. Kamihara, M. Hirano, and H. Hosono,
J. Phys. Soc. Japan {\bf 77}, 073701 (2008).

\bibitem{yildirim}
T. Yildirim,
arXiv:0804.2252 (2008).

\bibitem{ishibashi}
S. Ishibashi, K. Terakura, and H. Hosono,
J. Phys. Soc. Japan {\bf 77}, 053709 (2008).

\bibitem{yin}
Z.P. Yin, S. Lebegue, M.J. Han, B. Neal, S.Y. Savrasov, and W.E. Pickett,
arXiv:0804.3355 (2008).

\bibitem{mazin-2}
I.I. Mazin, M.D. Johannes, L. Boeri, K. Koepernik, and D.J. Singh,
Phys. Rev. B (in press).

\bibitem{vildosola}
V. Vildosola, L. Pourovskii, R. Arita, S. Biermann, and A. Georges,
arXiv:0806.3285 (2008).

\bibitem{yildirim-2}
T. Yildirim,
arXiv:0807.3936 (2008).

\bibitem{mcguire}
M.A. McGuire, A.D. Christianson, A.S. Sefat, B.C. Sales, M.D. Lumsden, R. Jin,
E.A. Payzant, D. Mandrus, Y. Luan, V. Keppens, V. Varadarajan, J.W. Brill,
R.P. Hermann, M.T. Sougrati, F. Grandjean, and G.J. Long,
arXiv:0806.3878 (2008).

\bibitem{mannella}
F. Bondino, E. Magnano, M. Malvestuto, F. Parmigiani, M.A. McGuire, A.S. Sefat,
B.C. Sales, R. Jin, D. Mandrus, E.W. Plummer, D.J. Singh and N. Mannella,
arXiv:0807.3781 (2008).

\bibitem{sf-note}
In a system with strong magnetic tendencies ordering will normally occur
at high temperature unless there is a large ${\bf q}$ space for fluctuations,
e.g. due to frustration, itinerant
electron physics or competing exchange interactions. It is difficult
to avoid ordering with large amplitude spin fluctuations if they are
they are strongly peaked, e.g. at the SDW nesting vector, or are subject
to a simple exchange model that favors a specific ordering pattern, such
as a nearest neighbor antiferromagnetic interaction on the square lattice
(with 3D interactions or magnetoelastic coupling as a present here).
Also, in the case where there is large phase space for fluctuations,
local magnetic correlations will be very short range. This does not
mean however that the fluctuations will necessarily be fast or only
at high energy and in fact near a quantum critical point there will be
slow fluctuations that extend to low energy.
One experimental signature of such fluctuations besides the magneto-structural
effects noted, will be via transport data and in particular data that show
nearness to a quantum critical point through non-Fermi liquid scalings.

\bibitem{moriya}
T. Moriya and K. Ueda,
Rep. Prog. Phys. {\bf 66}, 1299 (2003).

\bibitem{monthoux}
P. Monthoux and G.G. Lonzarich,
Phys. Rev. B {\bf 63}, 054529 (2001).

\bibitem{kuroki}
K. Kuroki, S. Onari, R. Arita, H. Usui, Y. Tanaka, H. Kotani, and H. Aoki,
arXiv:0803.3325 (2008).

\bibitem{mizuguchi}
Y. Mizuguchi, F. Tomioka, S. Tsuda, T. Yamaguchi, and Y. Takano,
arXiv:0807.4315 (2008).

\end{references}
\end{document}